\documentclass[conference, 10pt]{IEEEtran} 
\usepackage{epsfig,cite}
\usepackage{color}
\usepackage{amssymb, amsmath}

\def\A{\mathsf{A}}
\def\C{\mathsf{C}}
\def\G{\mathsf{G}}
\def\U{\mathsf{U}}
\def\var{\mathrm{Var}}
\def\editcolor{\color{black}}

\IEEEoverridecommandlockouts

\begin{document}


\title{RNA as a Nanoscale Data Transmission Medium: Error Analysis}
%

\author{
Andrew W.~Eckford$^1$, Taro Furubayashi$^2$, and Tadashi Nakano$^3$\\
$^1$Dept.~of Electrical Engineering and Computer Science, York University, Toronto, Canada M3J 1P3\\
$^{2,3}$Graduate School of Frontier Biosciences, Osaka University, Suita 565-0871, Japan\\
Email: $^1$aeckford@yorku.ca, $^2$fbayashi@fbs.osaka-u.ac.jp, $^3$tadasi.nakano@fbs.osaka-u.ac.jp%
\thanks{This work was supported in part by a grant from the Natural Sciences and Engineering Research Council (NSERC).}%
}

\maketitle

\begin{abstract}
RNA can be used as a high-density medium for data storage and transmission; however,
an important RNA process -- replication -- is noisy.
This paper presents an error analysis for RNA as a data transmission medium,
analyzing how deletion errors increase in a collection of replicated DNA strands over time.
\end{abstract}

\section{Introduction}
\label{sec:intro}

It is well known that genetic information is encoded at the nanoscale into deoxyribonucleic
acid (DNA) and ribonucleic acid (RNA). 
The extremely high density and stability of DNA/RNA-based data storage makes it a potential
solution for data storage and data transmission applications \cite{Goldman13}.
This scheme is an example of {\em molecular communication}, in which data is represented
and transmitted using the principles of chemistry \cite{Nakano-book}; this is a form of {\em nanonetworking} \cite{Bush-book,Akyildiz08}

Elsewhere, studies on {\em inscribed matter} as a data storage and transmission medium
has received recent attention. In \cite{rose04}, the energy-efficient properties of inscribed matter were considered as advantages for extremely long distance communication (e.g., between stars). 
DNA and RNA are examples of inscribed matter data storage and communication, with the data stored directly in the structure of the molecule. A competing technology, nanoscale laser-etched glass, recently demonstrated
the ability to record up to  360 terabytes on a small disc \cite{zhang16}, though this is far less than
the theoretical storage density of DNA/RNA. 

In our earlier work, we presented a scheme for 
inscribed-matter data transmission, in which data-bearing molecules were
released into the environment and propagated towards a receiver.
Using this scheme, large data strings could be conveyed from transmitter to receiver by 
dividing the data into packets, transmitting the data over the channel, and
reassembling the data at the receiver \cite{Furubayashi16}.
These schemes have been tested experimentally using RNA as the data-bearing molecule
\cite{Furubayashi15}.

Our results in this paper extend \cite{Furubayashi15,Furubayashi16}.
A significant issue with RNA data transmission is that RNA replication is noisy: although the
error rate is low, the
copies may not be perfect reproductions of the original.
Techniques in communication and information theory may be used to mitigate the errors: for example,
an error-correcting code may be used, but it is necessary to have a mathematical model for
the noise process in order to ensure that the code is reliable and efficient.

The contribution of this paper is to estimate the error rates for RNA as a nanoscale data transmission
medium, thereby permitting the design efficient error-correcting codes, and
ensuring the reliability of RNA-based communication systems.

\section{Model}

\subsection{Data transmission model}

In biological organisms, the information encoded in DNA and RNA often represents 
sequences of amino acids that are to be assembled into proteins. Generally, these sequences
are encoded in {\editcolor DNA or RNA; {\em transcription} is the process of reading DNA to produce RNA. Then, starting with RNA,} sequence is decoded to produce the correct protein, {\editcolor a process known as {\em translation}}. This, very briefly, is the {\em central dogma of molecular biology} \cite{central10}.

However, in this paper we are more interested in the ability of RNA to encode and transmit arbitrary data.
{\editcolor For our purposes, an RNA molecule is single-stranded, and stores data} 
represented by four nucleotides: 
adenine ($\A$), cytosine ($\C$), guanine ($\G$), and uracil ($\U$). These nucleotides are arranged in a sequence along an RNA strand, creating a 4-ary data string.
This data can be read (and replicated) by RNA polymerase, generating copies of the data string.
A single strand of RNA with $n$ nucleotides has $4^n$ possible strings, storing $\log_2 4^n = 2 n$ bits
of information.


As a very simple example, suppose we want to communicate the short text message ``Hello'' in RNA.
In the standard model of communication, there
are three key components of a communication system: the encoder, channel, and decoder (see, e.g., \cite[Figure 7.1]{cover-thomas}).
For RNA-based communication, these are described as follows.
\begin{itemize}
	\item {\em Encoder.} The message must be represented as a string of RNA nucleotides.
	Using ASCII encoding, each character in the five-letter message ``Hello'' takes a number in the range $[0,255]$. The text
	message is now given by the vector of ASCII indices
	\begin{equation}
		[72, 101, 108, 108, 111] .
	\end{equation}
	Since we have four nucleotides, it is convenient to express these numbers in base-4; 
	any integer in the range $[0,255]$ can be expressed with four base-4 digits. In our example,
	the above ASCII string maps to the base-4 vector
	\begin{equation}
		[1020, 1211, 1230, 1230, 1233]
	\end{equation}
	Let $\A = 0$, $\C = 1$, $\G = 2$, and $\U = 3$ in a base-4 representation of integers.
	Then the above base-4 vector can be mapped to the sequence of nucleotides
	\begin{equation}
		\label{eqn:RNA-sequence}
		\C\A\G\A \C\G\C\C \C\G\U\A \C\G\U\A \C\G\U\U .
	\end{equation}
	The sequence (\ref{eqn:RNA-sequence}) is then synthesized
	to a strand of RNA. Various commercial suppliers provide a service for 
	custom RNA synthesis.%
	\footnote{There may be restrictions on the sequences that can be generated with
	custom RNA synthesis. For example, certain sequences may lead
	to secondary structures; see \cite{kramer81}. For simplicity, we disregard these
	effects, though encoding schemes can be designed to avoid bad sequences.}
	For the purposes of this paper, we will assume that one strand of RNA with
	the exact sequence (\ref{eqn:RNA-sequence}) is available.
	
	\item {\em Channel.} The communication channel is the connection between transmitter
	and receiver. A key feature of RNA is that it can be replicated 
	using biological means: RNA polymerase ``reads'' the RNA molecule and produces a copy; it 
	is this replication mechanism that conveys the information, as one of the replicated strands will arrive at the receiver, for example after propagating via Brownian motion.
	However, the replication process introduces errors, and those errors can be passed down to 
	descendant copies of the RNA strand. In our example, suppose two replicas of the RNA
	strand are created, one of which contains a substitution error (highlighted in red):
	\begin{align}
		\nonumber
		\mathrm{Original} &\:\: \C\A\G\A \C\G\C\C \C\G\U\A \C\G\U\A \C\G\U\U \\
		\nonumber
		\mathrm{Replica \: 1} &\:\: \C\A\G\A \C\G\C\C \C\G\U\A \C\G\U\A \C\G\U\U \\
		\nonumber
		\mathrm{Replica \: 2} &\:\: \C\A\G\A \C\G\C\C \C{\color{red} \A}\U\A \C\G\U\A \C\G\U\U  
	\end{align}
	Now there are three replicas, one of which contains a single error.
	Each replica may, in turn, replicate; when Replica 2 replicates, the error is passed to all of
	its daughter replicas, in addition to any errors subsequently introduced in replication.
	
	\item {\em Decoder.} With many RNA sequences in the environment,
	the first RNA molecule to arrive at the receiver is sequenced to determine its
	information content. At the receiver, the decoding process is the inverse of the
	encoding process, converting the RNA sequence to a base-4 integer, and subsequently
	finding the correct ASCII characters. However, due to the noisy channel, there
	may be errors in the decoded sequence.

\end{itemize}
We emphasize that the example is given for illustration, and is
possibly not the most practical
way to express data in RNA.

As replication and error are key to our results, we give more specific details about the replication and reliable communication models below.


\subsection{Replication and error model}

Our RNA replication model uses RNA-dependent RNA polymerase,
which is used to replicate RNA in certain viruses.
{\editcolor A range of kinetic rates for this process are available in \cite{Powdrill11}.
However, this process is similar to transcription (defined above), for which good kinetic models
are available, e.g. those found in \cite{vonHippel98,Bai04}; these kinetic rates
are consistent with the ranges in \cite{Powdrill11}. However, the final
results should be robust to changes in model.}

Let $\mathbf{s} = [s_1,s_2,\ldots,s_n]$ represent a length-$n$ 
string of nucleotides encoded in RNA (i.e., the RNA data string),
where $s_i \in \{\A,\C,\G,\U\}$.
To ease our analysis we make several simplifying assumptions:
\begin{itemize}
	\item Transcription times are stochastic. The transcription time is independent
	for each letter $s_i$, has finite variance, and is identically distributed for the same nucleotide. (For example, if
	$s_i = s_j = \A$, then the transcription time for $s_i$ and $s_j$ are independent and 
	have the same distribution.)
	\item RNA does not degrade during the data transmission session. 
	\item Enough RNA polymerase and ATP is present to permit every RNA molecule to replicate simultaneously.
	\item All effects other than the replication of each individual nucleotide can be ignored.
\end{itemize}
Some of these assumptions can be relaxed. 

%

Let $t_i$ represent the transcription time for letter $s_i$. Then by the above assumptions, the total transcription time $T$ is
\begin{equation}
	T = \sum_{i=1}^n t_i .
\end{equation}
Since $t_i$ are independent and have finite variance, by the central limit theorem, 
the distribution of $T$ approaches the Gaussian distribution for large values of $n$.
Thus, the distribution of $T$ is dependent on its mean and variance:
\begin{align}
	E[T] &= n E[t_i] \\
	\label{eqn:String-1}
	&= n_\A E[t_\A] + n_\C E[t_\C] + n_\G E[t_\G] + n_\U E[t_\U]\\
	\var[T] &= n \var[t_i] \\
	\label{eqn:String-2}
	&= n_\A \var[t_\A] + n_\C \var[t_\C] + n_\G \var[t_\G] + n_\U \var[t_\U] ,
\end{align}
where $n_\A$ is the number of $\A$ letters in the string, and $t_\A$ represents the
time given that the string letter is $\A$ (similarly for $\C$, $\G$, $\U$). 


\subsection{Reliable communication model}

Replication of RNA introduces errors. Replication errors are 
insertions (i.e., a spurious new nucleotide that was not in the original sequence), 
deletions (i.e., a nucleotide removed from the sequence), {\editcolor or substitutions (i.e., 
the replacement of a nucleotide with a different nucleotide). Together, insertions and deletions are called {\em indels}. Recombination can also occur as part of the replication process, but we
do not consider recombination in this paper.}

The corruption of RNA transcription by indels is a good example of an {\em insertion-deletion channel}, which has long been studied by information theorists \cite{Diggavi01}.
Moreover, channels with substitution errors are commonly studied in information theory,
such as the binary symmetric channel (BSC); the combination of indel and substitution channels
has also been studied \cite{Davey01}. 
Information-theoretic analysis permits the construction of an error-correcting code, which could
recover data in spite of indel {\editcolor and substitution} errors. Error-correcting codes work by introducing redundancy into the information string; based on the rate of insertions and deletions, the minimum needed redundancy (and hence the highest {\em information rate}) is given by the Shannon capacity. However, in order to determine the Shannon capacity, we need to estimate the final
indel and substitution rates introduced by the channel when replicated RNA arrives at the receiver, which is what we do in this paper.

Error rates were estimated to be {\editcolor $2.3 \times 10^{-7}$ per nucleotide for insertions and deletions (indels), and $9.1 \times 10^{-6}$ for substitutions, in the Q$\beta$ bacteriophage \cite{Garcia12}.}
Using this model, in this paper we estimate the rate of {\editcolor insertions}, deletions, and substitutions of a randomly-selected RNA strand. 

\section{Results}

We simulate the RNA replication process to give insight on the ultimate probability of 
insertion,
deletion, and substitution in received RNA sequences.

We first consider insertions and deletions in isolation. In this case, we are particularly interested in two features of RNA data transmission systems: first,
we are interested in the average insertion or deletion rate under various conditions, as this is directly related to the Shannon capacity; and second, given that the reproduction rates of individual sequences vary according to their content (see equations (\ref{eqn:String-1}),(\ref{eqn:String-2})), the deletion probabilities may be different for different nucleotides. Owing to the complexity of these features, we study these probabilities in {\em Monte Carlo} simulations, which will guide future analytical work.

\begin{figure}[t!]
\begin{center}
\includegraphics[width=3.5in]{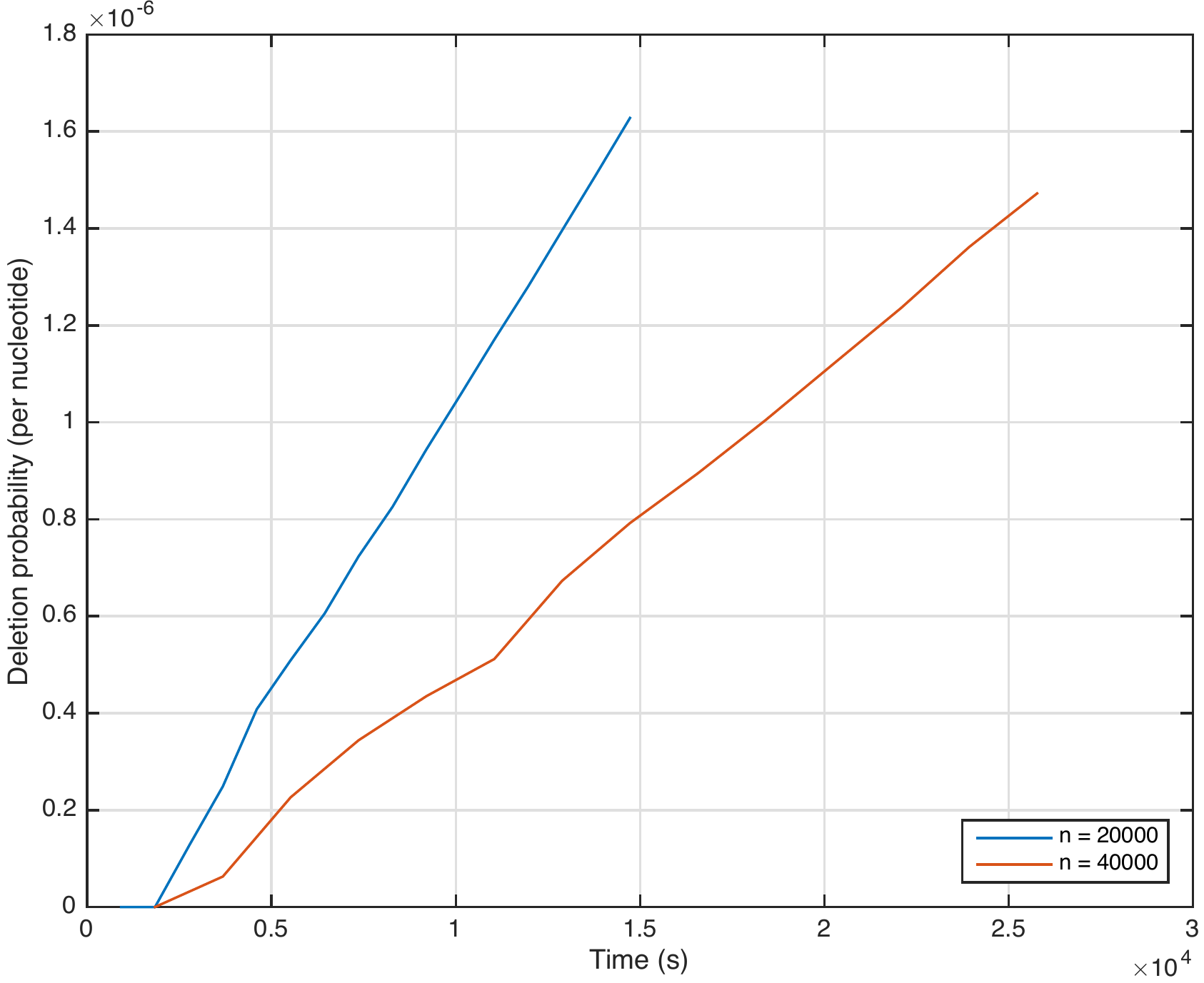}
\end{center}
\caption{\label{fig:AverageErasureProbability} Estimated deletion rate for a randomly selected strand of replicated RNA as a function of {\editcolor total transcription time} for $n = 20000$ and $n = 40000$. Replication times for each type of nucleotide ($\A$, $\C$, $\G$, $\U$) were assumed equal to 0.046 s. Initial deletion rate per nucleotide was $2.3 \times 10^{-7}$. Insertion rate should be the same.}
\end{figure}

Using a reaction rate of $k = 22$ s$^{-1}$ reactions per nucleotide (cf. \cite[Eqn (3)]{Bai04}; {\editcolor see also \cite{Powdrill11}}), we obtain
\begin{align}
	E[t_i] &= 1/k = 0.046 \: \mathrm{s} \\
	\var[t_i] &= 1/k^2 = 2.1 \times 10^{-3} \: \mathrm{s}^2
\end{align}
We use the provided indel rate of $2.3 \times 10^{-7}$ insertions and deletions per nucleotide.
Using {\em Monte Carlo} simulations, we can estimate the deletion probability as a function of time, as shown in Figure \ref{fig:AverageErasureProbability}; we give results for two RNA string lengths ($n = 20000$ and $n = 40000$). (As the insertion probability is equal, the resulting insertion probability should be the same.)

\begin{figure}[t!]
\begin{center}
\includegraphics[width=3.5in]{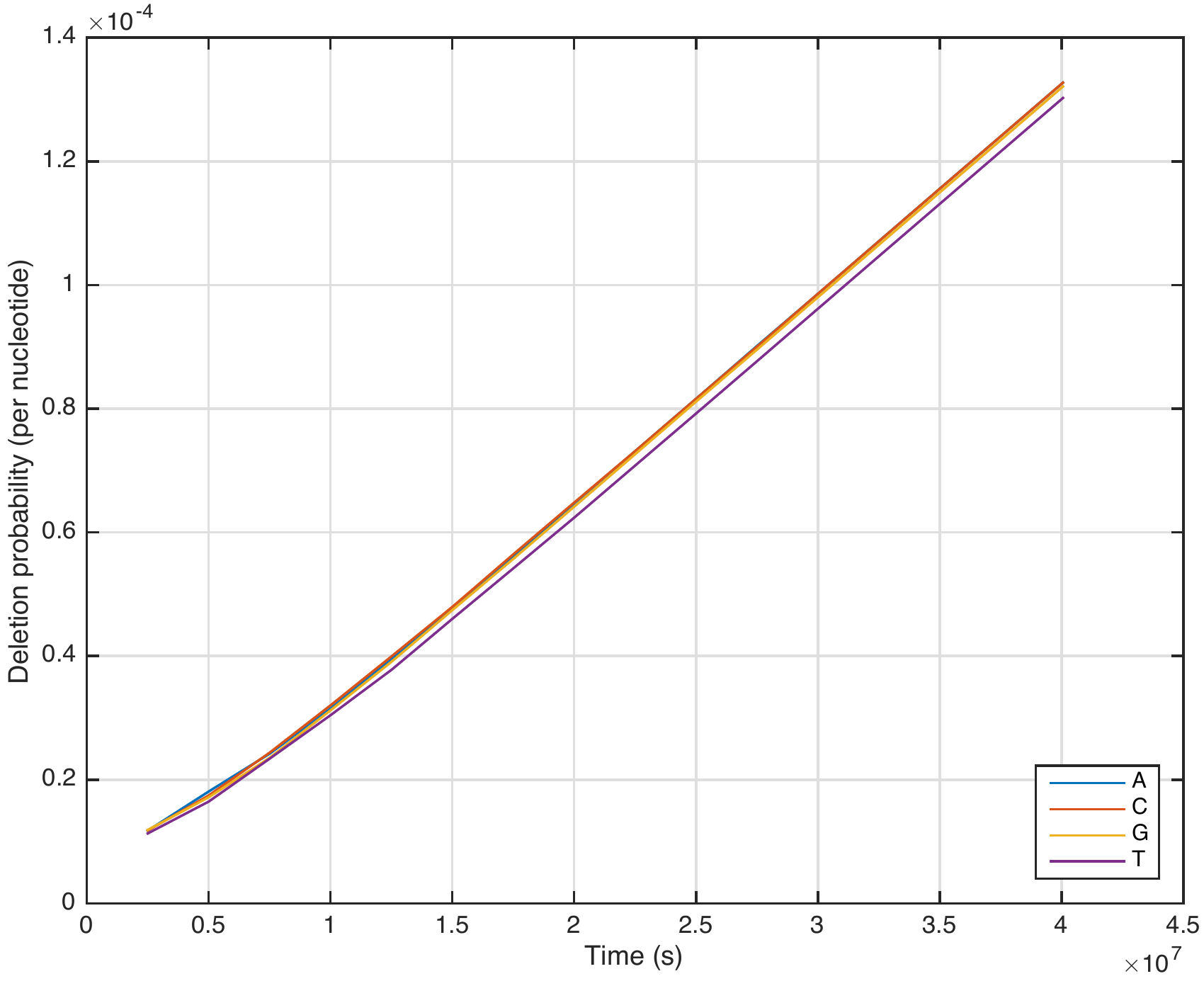}
\end{center}
\caption{\label{fig:EachLetterErasureProbability} Estimated deletion rate for a randomly selected strand of replicated RNA as a function of {\editcolor total transcription time} for $n = 100000$, and different replication rates per nucleotide (see equations (\ref{eqn:DifferentRates1})-(\ref{eqn:DifferentRates2})). Initial deletion rate per nucleotide was $1.7 \times 10^{-5}$. The substantially longer time, compared with Figure \ref{fig:AverageErasureProbability}, is due to the lower replication rate for $\A$.}
\end{figure}

To test the effect of different replication rates on erasure rate, we ran another simulation with
\begin{align}
	\label{eqn:DifferentRates1}
	E[t_\A] &= 0.001 \: \mathrm{s} \\
	\label{eqn:DifferentRates2}
	E[t_\C] = E[t_\G] = E[t_\U] &= 0.046 \: \mathrm{s}
\end{align}
with variances equal to the squares of the expected values, as before. Here, to emphasize the effect of deletions, we increased the deletion probability to $1.7 \times 10^{-5}$ and increased the block length to $n = 100000$. Here, $E[t_\A]$
was chosen arbitrarily, to create a large contrast in reproduction time between $\A$ and the
other nucleobases; strings started out with nearly equal numbers of each nucleobase. These results are shown in Figure \ref{fig:EachLetterErasureProbability}.

Our observations on the results:
\begin{itemize}
	\item From Figure \ref{fig:AverageErasureProbability}, erasure probability grows linearly with time. Different string length has a large effect,
	likely because shorter strings replicate faster. The dominant effect in determining deletion
	probability appears to be {\em generation}, i.e., how many replications have taken place since the beginning.
	\item Surprisingly, from Figure \ref{fig:EachLetterErasureProbability}, we see that there is essentially no effect on erasure
	rate as a result of different nucleotide replication rates.
\end{itemize}
These results set up an interesting tradeoff: the longer the delay, the more replicated packets are produced (and hence the more likely one arrives quickly); however, this is at the cost of increasing deletions.

\begin{figure}[t!]
\begin{center}
\includegraphics[width=3.5in]{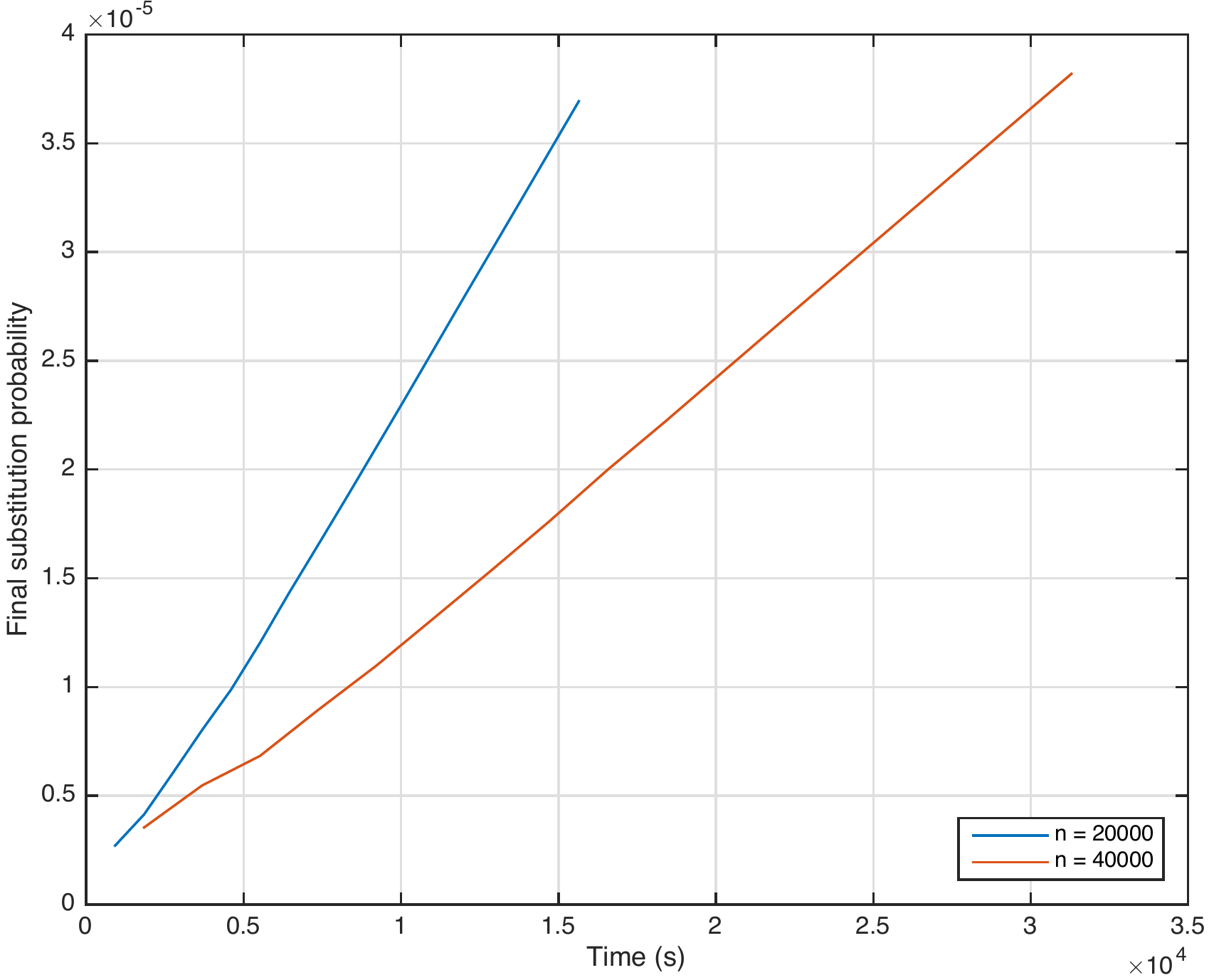}
\end{center}
\caption{\label{fig:AverageSubstitutionProbability} Estimated substitution rate for a randomly selected strand of replicated RNA as a function of {\editcolor total transcription time} for $n = 20000$ and $n = 40000$. Replication times for each type of nucleotide ($\A$, $\C$, $\G$, $\U$) were assumed equal to 0.046 s. Initial substitution rate per nucleotide was $9.1 \times 10^{-6}$. Indel rate was the same as in Figures 1 and 2.}
\end{figure}

In Figure \ref{fig:AverageSubstitutionProbability}, we use similar {\em Monte Carlo} simulations to estimate the rate of substitution
errors. In these simulations, we use the provided substitution rate of $9.1 \times 10^{-6}$; indels are also included in the simulation at the rates used previously. We make two observations on these results: first, that the behaviour of substitution errors as a function of time is similar to the behaviour of indel errors; and second, that the substitution error rate is clearly much higher, owing to the higher rate in each replication.

\section{Discussion and Conclusion}

Results given in this paper give an idea of how to design reliable RNA data storage systems.
While the decoded sequences contain errors, as a result of the unreliability of RNA replication,
these can be mitigated by error-correcting codes. However,
knowledge of the indel and substitution rates are required to design efficient error correcting codes: 
these quantities allow the designer to estimate the needed redundancy (in the sense of Shannon capacity) in order to recover a given information string. 
For example, considering only indels, if the probability of an indel error is $p_D$, then at least $p_D / (1-p_D)$ redundant nucleotides are required per information-bearing nucleotide. This is a lower bound, called the erasure channel bound, which is achieved if an oracle informs you of the locations of every indel (but not the identity of any deleted nucleotide). In practice, more redundancy would be needed. Estimation of the Shannon capacity from the insertion-deletion-substitution rate 
is more complicated, and will be considered in future work.


\bibliographystyle{ieeetr}


\end{document}